\documentclass[12pt]{article}

\usepackage{graphics}
\newcommand{\comment}[1]{}
\newcommand{\bea}{\begin{eqnarray}}
\newcommand{\eea}{\end{eqnarray}}
\newcommand{\be}{\begin{equation}}
\newcommand{\ee}{\end{equation}}

\begin{document}

\begin{center}
{\Large \bf \boldmath T-odd observables from anomalous 
$tbW$ couplings in single-top production \\ \vskip .2cm
at an $ep$ collider}

\bigskip
\bigskip
\bigskip

{\large Saurabh D. Rindani}

\bigskip

{\it Theoretical Physics Division, Physical Research Laboratory,\\ 
\vskip .1cm
Navrangpura, Ahmedabad 380009, India}

\end{center}

\bigskip
\bigskip
\centerline{\large Abstract}

\bigskip
We investigate the possibility that an imaginary 
anomalous $tbW$
coupling can be measured in the process $e^-p \to \nu_e \overline t X$
by means of T-odd observables.
One such observable considered here is the polarization of the top 
antiquark transverse
to the production plane. The other is a T-odd 
correlation constructed out of observable
momenta when the top quark decays leptonically. Both these T-odd
observables are shown to be proportional to the imaginary part of only 
one of the $tbW$ anomalous couplings, the other couplings giving
either vanishing or negligible contribution. This imaginary part could
signal either a CP-odd coupling, or an absorptive part in the effective
coupling, or both. We estimate the 1-$\sigma$ limits that might be
derived in the case of each of these observables for a collider with
a proton energy of 7 TeV and an electron energy of 60 GeV and also 
in the case of a higher electron energy of 150 GeV.

\section{Introduction}

The Large Hadron Collider (LHC), a proton-proton collider, 
 has seen many successes. With an
enhanced luminosity, as in the future runs such as the High-Luminosity
version (HL-LHC), results with high precision are anticipated.
Another option for the future which has been considered is an
electron-hadron collider, which will have some advantages over the LHC.
The Large Hadron electron Collider (LHeC) would thus be a useful
companion machine for the HL-LHC, allowing full exploitation of the 
data and significantly extending its reach 
\cite{CDR, LHeCStudyGroup:2012zhm, haosun}. 
The main advantage of the
LHeC would be because of the cleaner environment of an electron beam
over a proton beam, 
with suppressed backgrounds from strong interactions. It would also be 
free from issues like pile-ups and 
multiple interactions. In addition, the backward and forward scattering 
can be distinguished due to the asymmetric initial states.
Suggestions have been made for an electron beam with
an energy of 60 GeV with the proton beam of energy 7 TeV available at
the LHC. There also exist proposals for Future Circular Collider which
may have an electron-hadron version apart from a hadron-hadron version.
This version of $ep$ collider would operate at even higher energies.

Apart from investigating with great accuracy the parameters of the
standard model (SM), the future machines will also strive to discover
new physics -- either in terms of discovery limits on new particles, or
extending limits on existing couplings of the SM particles. 

One such investigation carried out was related to the measurement of
anomalous $tbW$ couplings 
through the cross section and kinematic distributions
of final-state particles in the single-antitop process $e^-p \to \nu_e 
\overline tX$ at an electron-proton collider \cite{Dutta:2013mva}.
The process occurs through the exchange of a virtual $W$ between an 
$e^-$ and a $\overline b$ quark within the proton, producing 
a $\overline t$ 
through an anomalous $tbW$ vertex, as shown in the diagram in Fig. 1.
\begin{figure}[h]
\begin{center}
\includegraphics{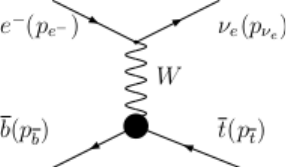}
\caption{ The Feynman diagram contributing to  $e^-\overline b \to \nu_e \overline t$ production. 
The effective $tbW$ vertex is shown as a filled circle.}\label{feyngraph1}
\end{center}
\end{figure}
The $W$ helicity fraction in $\overline t$ decay 
has also been suggested as a cue to the
measurement of the anomalous couplings \cite{Dutta:2013mva}. 
Longitudinal top polarization in an $e^+p$  process 
has been discussed by Atag and Sahin \cite{Atag:2006by}
as a means to study anomalous $tbW$ couplings. 

We concentrate here on observables in the above single-antitop
production process which are odd under naive time 
reversal (T). By this term we mean an operation limited to
reversing the signs of momenta and spins of all particles. By contrast,
genuine time reversal also involves interchange of initial and
final states, and is therefore a transformation difficult to achieve in
 actual practice. 
At tree level, unitarity implies that the forward and backward
amplitudes have the same phase, and therefore the time reversal
violation, if present,
would be genuine. It would therefore occur
only in the presence of CP violation, on account of the CPT theorem.
On the other hand if the amplitude has an absorptive part arising from
rescattering or loop diagrams, characterized by a nonzero imaginary 
part for an anomalous effective coupling, 
the forward and backward amplitudes are
not the same, and the subsequent T violation would not be a genuine one.
In that case the CPT theorem does not necessarily imply CP violation.

We study in the process $e^- \overline b \to \nu_e \overline t$ 
two observables which are T odd: (1) the polarization of the $\overline
t$ perpendicular to the production plane (the transverse
polarization) and (2) a combination of angular variables of the
$\overline t$ and the charged lepton arising in $\overline t$ decay,
which corresponds to the angular part of the three-vector triple product
$\vec p_{e^-}\times \vec p_{\overline t} \cdot \vec p_{\ell^-}$.
These are studied in the presence of possible anomalous $tbW$ couplings
in an effective Lagrangian. It is customary to consider an effective
Lagrangian arising in what is termed as an effective field theory, 
where the SM Lagrangian is
supplemented by higher dimensional terms which are suppressed by inverse
powers of a cut-off scale, a scale up to which the effective field
theory is valid. However, here we do not make any such specific
assumption about scales, but simply write an effective Lagrangian
permitted by Lorentz invariance.

\section{\boldmath Anomalous $tbW$ couplings: CP and T properties}
The Lagrangian for effective $tbW$ couplings may be written as
\begin{equation}\label{lag}
\begin{array}{rcl}
{\cal L}_{Wtb}& =& \displaystyle \frac{g}{\sqrt{2}} \left[ 
W_\mu \overline t \gamma^\mu (V_{tb} f_{1L} P_L + f_{1R} P_R)b \right.
 \\
 &&\left. \displaystyle  -\frac{1}{2 m_W} W_{\mu\nu}\overline t \sigma^{\mu\nu}
(f_{2L} P_L + f_{2R} P_R)b  \right] + {\rm H.c.}
\end{array}
\end{equation}
Here, $g$ is the semi-weak coupling, $P_{L,R}$ are the left (right) chirality projection matrices, and
$W_{\mu\nu} = \partial_\mu W_\nu - \partial_\nu W_\mu$.
We will later assume $V_{tb} = 1$ for simplicity. In the SM at tree
level, $f_{1L}=1$, and the remaining $f_i$ are vanishing.

In general, the anomalous couplings 
$f_{1R}$, $f_{2L}$ and $f_{2R}$ can be complex. The presence of an
imaginary part can signal either an absorptive part in the form factor
or
violation of CP or both. Whether the coupling is CP violating or not can
be checked 
by a comparison of the phases in the conjugate processes involving the
$t$ and the $\overline t$. The CP-violating phases would be opposite in
sign to each other in these conjugate processes. 
The absorptive phase, however, would be the
same in both these processes. A discussion of these issues may be found
in \cite{bern,Antipin:2008zx,sdrPankajPhases} and is summarized below.
Anomalous couplings, including the absorptive parts, have been
evaluated in certain theoretical scenarios in ref. \cite{bern}.

We can write the most general 
effective vertices up to mass dimension 5 for the four distinct 
processes of 
$t$ decay, $\overline t$ decay, $t$ production and $\overline t$ 
production as 
 \be\label{tdecay}
 V_{t\to bW^+} =- \frac{g}{\sqrt{2}} V_{tb} [ \gamma^\mu 
        (f_{1L} P_L + f_{1R} P_R ) 
        -i(\sigma^{\mu\nu}q_\nu/m_W)  (f_{2L} P_L + f_{2R}
        P_R ) ],
\ee
\be
 \label{tbardecay}
 V_{\bar t\to \bar bW^-} = -\frac{g}{\sqrt{2}} V_{tb}^*[\gamma^\mu 
        (\bar f_{1L} P_L + \bar f_{1R} P_R ) 
        -i(\sigma^{\mu\nu}q_\nu/m_W) (\bar f_{2L} P_L + \bar f_{2R}
        P_R )],
\ee
\be
 \label{tprod}
 V_{b^*\to tW^-} = - \frac{g}{\sqrt{2}} V_{tb}^* [ \gamma^\mu 
        (f_{1L}^* P_L + f_{1R}^* P_R ) 
        -i(\sigma^{\mu\nu}q_\nu/m_W) (f_{2L}^* P_R +
        f_{2R}^* P_L ) ],
\ee
\be
 \label{tbarprod}
 V_{\bar b^*\to \bar tW^+} = - \frac{g}{\sqrt{2}} V_{tb} [ \gamma^\mu 
        (\bar f_{1L}^* P_L + \bar f_{1R}^* P_R ) 
        -i(\sigma^{\mu\nu}q_\nu/m_W) (\bar f_{2L}^* P_R +
        \bar f_{2R}^* P_L )],
 \ee
where $f_{1L}$, $f_{1R}$, $f_{2L}$, $f_{2R}$,
$\bar f_{1L}$, $\bar f_{1R}$, $\bar f_{2L}$, and $\bar f_{2R}$ are form
factors, $P_L$, $P_R$ are left-chiral and right-chiral projection
matrices, and $q$
represents the $W^+$ or $W^-$  momentum, as applicable in each case.
In the SM, at tree level, $f_{1L}=\bar f_{1L} = 1$, and all other form
factors vanish. New physics effects would result in deviations of
$f_{1L}$ and  $\bar f_{1L}$ from unity, and nonzero values for other
form factors.

At tree level, when there are no absorptive parts in the
relevant amplitudes giving rise to the form factors, the following
relations would  be obeyed: 
 \be\label{noabspart}
 \begin{array}{rccl}
 f_{1L}^* = \bar f_{1L};& f_{1R}^* = \bar f_{1R}; & f_{2L}^* = \bar
 f_{2R}; & f_{2R}^* = \bar f_{2L}.
 \end{array}
 \ee
This can be seen to follow from the fact that in the absence of
absorptive parts, the effective
Lagrangian is Hermitian. Thus, the Hermiticity of the Lagrangian
of eq. (\ref{lag})
from which the amplitudes
(\ref{tdecay})-(\ref{tbarprod}) may be derived implies the relations 
(\ref{noabspart}).
On the other hand, if CP is conserved, the relations obeyed by the form
factors are
\be\label{CP}
\begin{array}{rccl}
f_{1L} = \bar f_{1L};& f_{1R} = \bar f_{1R}; & f_{2L} = \bar
f_{2R}; & f_{2R} = \bar f_{2L}.
\end{array}
\ee

The phases of the form factors can be expressed as sums and 
differences of two phases, one corresponding to a nonzero absorptive
part, and another corresponding to CP nonconservation:
 \be\label{phases}
 \begin{array}{rl}
 f_{1L,R} = \vert f_{1L,R} \vert \exp{(i\alpha_{1L,R} + i\delta_{1L,R})};&
 \bar f_{1L,R} = \vert f_{1L,R} \vert \exp{(i\alpha_{1L,R} -
 i\delta_{1L,R})};\\
 f_{2L,R} = \vert f_{2L,R} \vert \exp{(i\alpha_{2L,R} + i\delta_{2L,R})};&
 \bar f_{2R,L} = \vert f_{2L,R} \vert \exp{(i\alpha_{2L,R} - i\delta_{2L,R})},\\
 \end{array}
 \ee
where, for convenience, magnitudes of the couplings are assumed to be
equal in appropriate pairs.
In case there are no absorptive parts ($\alpha_i = 0$), we get the
relations (\ref{noabspart}) as a special case. In case of CP
conservation, we get the relations (\ref{CP}) as a special case. In what
follows, we will deal with a more general case when both CP violation
and absorptive parts are present. 
However, even though the couplings which contribute to
$\overline t$ production are $\overline f_i$, in what follows, we will
simply use $f_i$, consistent with the notation of, for example, 
\cite{Dutta:2013mva}.

\section{T-odd observables}
We investigate two representative T-odd observables. The first one is 
the polarization of the top antiquark transverse to the
production plane. 
The polarization of top quarks may be measured using the kinematic
distributions of the decay particles, without contamination
from the hadronization process, since the quark being heavy 
is expected to decay
before hadronization. 

The polarization can be measured experimentally from the decay
distribution of one of the decay products in the rest frame of the
top antiquark.
The angular distribution of a decay product $f$ for an antitop
 ensemble has the form
 \begin{equation}
 \frac{1}{\Gamma_f}\frac{\textrm{d}\Gamma_f}{\textrm{d} \cos \theta _f}
=\frac{1}{2}(1+\kappa _f P_{\overline t} \cos \theta _f).
 \label{topdecaywidth}
 \end{equation}
 Here $\theta_f$ is the angle between the momentum of fermion $f$ 
and the antitop spin
vector in the antitop rest frame and $P_{\overline t}$
is the degree of polarization of the antitop ensemble.
$\Gamma_f$ is the partial decay width and $\kappa_f$ is the spin
analyzing power of $f$. For a charged lepton $f \equiv \ell^-$ 
the analyzing power is $\kappa_{\ell^-} = -1$ at tree level. 
We do not discuss the details of such a polarization 
measurement. However, it is appropriate to note that a measurement of
the antitop polarization using the  angular distribution of
$\ell^-$ will not be affected by the contribution of an anomalous 
$tbW$ coupling in the top decay process to linear order, as shown in
earlier work \cite{theorem}.

The other T-odd observable involves using the momentum
of an antitop decay particle and constructing a triple vector product
of its three-momentum with the incoming $e^-$ momentum and $\overline t$
momentum. Since three-vectors change sign under T, such a triple product
would be odd under T. As explained later, we will use only the factors
involving angular variables and not the whole triple product.

Both these observables, being T odd, depend on the existence of an
absorptive amplitude, as discussed earlier. In an approach with 
 an effective Lagrangian
like the one in eq. (\ref{lag}), this implies that the observable will be
proportional to the imaginary part of an effective coupling. In our
case, it turns out that Im~$f_{2L}$ is the only one which contributes
significantly.
 
In order to calculate the T-odd observables,
we first obtain analytic expressions for the
production spin density matrix elements.  
For calculating helicity amplitudes and
density matrix elements, 
we have made use of the algebraic manipulation software FORM \cite{verm}
 as well as well the procedure outlined in \cite{VegaWudka}.
We also calculated the transverse polarization of the top antiquark
using FORM.
We then evaluate numerical expressions for the scenarios with electron
energy $E_e = 60$ GeV 
and a proton energy $E_p = 7$ TeV.
We shall also look at the possibility of an electron  beam of 
higher energy of
$E_e = 150$ GeV being available, with the same proton energy.

\section{Spin density matrix for antitop production}
We give below  the expressions  for $\rho_{ij}$, the
density matrix elements for the production process $e^-\overline  b \to
\nu_e\overline t $ 
where $i,j = 1,2$ refer
respectively to the helicities $+1/2$ and $-1/2$ of the $\overline t$. 
The expressions 
are listed separately for the cases where
one anomalous coupling is assumed nonzero at a time, the remaining assumed
to be vanishing.

\smallskip

\noindent Case (i): $f_{1R} \neq 0$, $f_{2L}=f_{2R}=0$.
\begin{equation}
\rho_{11} = 
\frac{g^4}{16}|\Delta_W|^2 
(\hat s - m_b^2)( \hat s - m_t^2) 
((1-\cos\theta_{\overline t}^{\rm
cm})^2 + (m_b^2/\hat s) \sin^2\theta_{\overline t}^{\rm cm})
\end{equation}
\begin{equation}
\begin{array}{rcl}
\rho_{22}\!\!\! &=& \!\!\!
\displaystyle
\frac{g^4}{16}|\Delta_W|^2 
(\hat s - m_b^2)( \hat s - m_t^2) 
 [
  {\rm Im} f_{1R}^2 +
({\rm Re} f_{1R} - (m_bm_t/\hat s)
(1 + \cos\theta_{\overline t}^{\rm cm}))^2 \\&& +
\displaystyle
(m_t^2/\hat s) \sin^2\theta_{\overline t}^{\rm cm}]
\end{array}
\end{equation}
\begin{equation}
\begin{array}{rcl}
\rho_{12} \;=\;   \rho_{21}^*\!\!&=\!\! &
\displaystyle
\frac{g^4}{16}|\Delta_W|^2 
(\hat s - m_b^2) (\hat  s - m_t^2) \sin\theta_{\overline t}^{\rm cm} [ (m_b^2/\hat s)
(m_t/\sqrt{\hat s}) (1 + \cos\theta_{\overline t}^{\rm cm}) \\&&  - (m_b/\sqrt{\hat s}) 
({\rm Re} f_{1R} + i {\rm Im} f_{1R}) 
+ (m_t/\sqrt{\hat s})(1-\cos\theta_{\overline t}^{\rm cm})]
\end{array}
\end{equation}
\comment{
\begin{equation}
\begin{array}{rcl}
\rho_{21} &&= 
\displaystyle
\frac{g^4}{16}|\Delta_W|^2 
(\hat s - m_b^2) (\hat  s - m_t^2) \sin\theta_{\overline t}^{\rm cm} [ (m_b^2/\hat s)
(m_t/\sqrt{\hat s}) (1 + \cos\theta_{\overline t}^{\rm cm})\\&&   - (m_b/\sqrt{\hat s}) 
({\rm Re} f_{1R} - i {\rm Im} f_{1R}) 
+ (m_t/\sqrt{\hat s})(1-\cos\theta_{\overline t}^{\rm cm})]
\end{array}
\end{equation}
} 
In the above equations, $\sqrt{\hat s}$ is the $e^-b$ energy in the
centre-of-mass (cm)
frame of the pair, $\theta_{t}^{\rm cm}$ is the angle made by
the $\bar t$ momentum with the $e^-$ momentum, 
and $\Delta_W$ is the $W$ propagator factor
\begin{equation}
\Delta_W = \frac{1}{[( p_{e^-} - p_{\nu_e})^2 - m_W^2]}.
\end{equation}
For a $\overline b$ quark carrying a momentum fraction $x$ of the proton
energy, $\hat s \approx 4xE_eE_p$. 

\smallskip

\noindent Case (ii): $f_{2L} \neq 0$, $f_{1R}=f_{2R}=0$.
      \begin{equation}\rho_{11} =  \frac{g^4}{32}|\Delta_W|^2 
(\hat s-m_b^2) (\hat s-m_t^2)
 \left[(1 - \cos\theta_{t}^{\rm cm})^2 +
           (m_b^2/{\hat s})\sin^2\theta_{t}^{\rm cm}  \right]  \end{equation}
      \begin{equation}
\rho_{22} = \displaystyle 
\frac{g^4}{32}|\Delta_W|^2 
|F_{2L}|^2 \frac{(\hat s-m_b^2)(\hat s-m_t^2) m_t^2}{\hat s}
     	\left[\sin^2\theta_{t}^{\rm cm} + (m_b^2/{\hat s})
          (1+\cos\theta_{t}^{\rm cm})^2 
     	   \right]
\end{equation}
      \begin{equation}
\begin{array}{rcl}
      \rho_{12}\;\, = \;\,
        \rho_{21}^* &=&
\displaystyle
\frac{g^4}{32}|\Delta_W|^2 
  F_{2L}\frac{(\hat s-m_b^2)(\hat s-m_t^2) 
           m_t}{\sqrt{\hat s}}\sin\theta_{t}^{\rm cm} 
\\&& \times \displaystyle \left[2  -(1- m_b^2/{\hat s}) (1 
 +\cos\theta_{t}^{\rm cm}) \right]
\end{array}
\end{equation}
\comment{
      \begin{equation}\rho_{21} =  
\frac{g^4}{32}|\Delta_W|^2 
F_{2L}^* \frac{(\hat s-m_b^2)(\hat s-m_t^2)
          m_t}{\sqrt{\hat s}}\sin\theta_{t}^{\rm cm} 
\left[ 2-(1-m_b^2/{\hat s}) (1+\cos\theta_{t}^{\rm cm}) \right]
\end{equation}
} 
where
 \begin{equation}	
F_{2L} = 1 + ({\rm Re} f_{2L} + i{\rm Im} f_{2L})\frac{(\hat s-m_t^2)}
{m_tm_W}.
\end{equation}

\smallskip

\noindent Case (iii): $f_{2R} \neq 0$, $f_{1R}=f_{2L}=0$.
\begin{equation}
\rho_{11} = 
\frac{g^4}{16}|\Delta_W|^2 
            (\hat s-m_b^2)(\hat s-m_t^2)[(1-\cos\theta_{\overline t}^{\rm cm})^2
     + |F_{2R}|^2 (m_b^2/\hat s) \sin^2\theta_{\overline t}^{\rm cm}],
\end{equation}
\begin{equation}
\rho_{22} = 
\frac{g^4}{16}|\Delta_W|^2 
(\hat s - m_b^2)(\hat s - m_t^2)\frac{m_t^2}{\hat s}
[\sin^2\theta_{t}^{\rm cm} + (m_b^2/\hat s)|F_{2R}|^2
(1+\cos\theta_{\overline t}^{\rm cm})^2 ],
\end{equation}
\begin{equation}
\begin{array}{rcl}
      \rho_{12}\;\, = \;\,
	\rho_{21} &=&
\displaystyle
\frac{g^4}{16}|\Delta_W|^2 
\frac{\hat s- m_b^2)(\hat s - m_t^2)m_t}{\sqrt{\hat s}}  
     \sin\theta_{\overline t}^{\rm cm}[(1 - \cos\theta_{\overline t}^{\rm} )\\&&
\displaystyle
          + \frac{m_b^2}{\hat s}|F_{2R}|^2  (1 +
     	\cos\theta_{\overline t}^{\rm cm})],
\end{array}
\end{equation}
where
 \begin{equation}	
F_{2R} = 1 + ({\rm Re} f_{2R} + i{\rm Im} f_{2R})\frac{(\hat s-m_b^2)}
{m_bm_W}.
\end{equation}

\comment{ Including the contribution of $f_{2L}$
      \begin{equation}\rho_{11} =  \frac{g^4}{32}|\Delta_W|^2 
(\hat s-m_b^2) (\hat s-m_t^2)
 \left[(1 - \cos\theta_{t}^{\rm cm})^2 +
           (m_b^2/{\hat s})\sin^2\theta_{t}^{\rm cm}  \right]  \end{equation}
      \begin{equation}
\rho_{22} = \displaystyle 
\frac{g^4}{32}|\Delta_W|^2 
|F_{2L}|^2 \frac{(\hat s-m_b^2)(\hat s-m_t^2) m_t^2}{\hat s}
     	\left[\sin^2\theta_{t}^{\rm cm} + (m_b^2/{\hat s})
          (1+\cos\theta_{t}^{\rm cm})^2 
     	   \right]
\end{equation}
      \begin{equation}
\rho_{12} = 
\frac{g^4}{32}|\Delta_W|^2 
  F_{2L}\frac{(\hat s-m_b^2)(\hat s-m_t^2) 
           m_t}{\sqrt{\hat s}}\sin\theta_{t}^{\rm cm} 
 \left[2-(1- m_b^2/{\hat s}) (1+\cos\theta_{t}^{\rm cm}) \right]
\end{equation}
      \begin{equation}\rho_{21} =  
\frac{g^4}{32}|\Delta_W|^2 
F_{2L}^* \frac{(\hat s-m_b^2)(\hat s-m_t^2)
          m_t}{\sqrt{\hat s}}\sin\theta_{t}^{\rm cm} 
\left[ 2-(1-m_b^2/{\hat s}) (1+\cos\theta_{t}^{\rm cm}) \right]
\end{equation}
} 

The expression for the production cross section in terms of the elements
 of $\rho$ is
\begin{equation}\label{sigmaSM}
\sigma =
\int dx\, \overline b(x) \int 
d\cos\theta_{\overline t}^{\rm cm} \,
\frac{|\vec p_{\,\overline t}^{\,\rm cm}|}{32 \pi \hat s E_e^{\rm cm}}
(\rho_{11} + \rho_{22}),
\end{equation}
where the superscript ``cm'' refers to the $e^- \overline b$ cm frame. 
$\overline b(x)$ is the parton distribution function of the $b$
antiquark in the proton, and where we employ the five-flavour scheme. 

\section{\boldmath Transverse polarization of $\overline t$}
The transverse antitop polarization, that is, the polarization transverse to
the antitop production plane, arises for nonzero Im~$f_{2L}$. 
It is in fact the asymmetry 
\begin{equation}
A_{\rm pol.}=
\frac{
\sigma(\vec p_{e^-} \times \vec
p_{\overline t} \cdot \vec s_{\overline t} >0) 
- 
\sigma(\vec p_{e^-} \times \vec
p_{\overline t} \cdot \vec s_{\overline t} <0)
}{ 
\sigma(\vec p_{e^-} \times \vec
p_{\overline t} \cdot \vec s_{\overline t} >0) 
+ 
\sigma(\vec p_{e^-} \times \vec
p_{\overline t} \cdot \vec s_{\overline t} <0)
},
\end{equation}
where $\vec s_{\overline t}$ refers to the spin of the antitop. 
We have evaluated the asymmetry to linear order in the anomalous 
coupling.

The numerator of this asymmetry, which may be called 
 the polarized cross section $\sigma_{\rm pol.}$,
is an integral over the phase space of the squared matrix element for
the production of an antitop with polarization perpendicular to the
production plane. That is, choosing the electron direction as the $z$
axis and the antitop direction in the $xz$ plane, it is the polarization
along the $y$ direction. The spin vector of the antitop is chosen 
along the $y$ direction to evaluate the relevant squared matrix element.

  The result for the numerator is
\begin{equation}
\sigma_{\rm pol.} = 
\int dx\, \overline b(x) \int 
d\cos\theta_{\overline t}^{\rm cm} \,
\frac{|\vec p_{\,\overline t}^{\,\rm cm}|}{32 \pi \hat s E_e^{\rm cm}}
[-i(\rho_{12}-\rho_{21})],
\end{equation}
where the quantity in square brackets is just $2 {\rm Im}~\rho_{12}$.
Thus, for the asymmetry to be nonzero, the off-diagonal density matrix
element $\rho_{12}$ has to have an imaginary part.
The denominator, which is the unpolarized cross section,
 is the same as $\sigma_{SM}$ given in eq. (\ref{sigmaSM}) above, which we
repeat for clarity:
\begin{equation}
\sigma_{\rm unpol.} =
\int dx\, \overline b(x) \int 
d\cos\theta_{\overline t}^{\rm cm} \,
\frac{|\vec p_{\,\overline t}^{\,\rm cm}|}{32 \pi \hat s E_e^{\rm cm}}
(\rho_{11} + \rho_{22}),
\end{equation}
We can now substitute the expressions we obtained for $\rho_{ij}$ for 
the various cases to obtain the asymmetries in those cases.

The most interesting case is Case (ii).
With only the anomalous coupling $f_{2L}$ taken as non-vanishing, the
result is
\begin{equation}
\begin{array}{lcl}
\displaystyle
\sigma_{\rm pol.}
& = &\displaystyle {\rm Im}\,f_{2L}\frac{g^4}{16}
\int dx\, \overline b(x) 
\int d\cos\theta_{\overline t}^{\rm cm} \,
\frac{|\vec p_{\,\overline t}^{\,\rm cm}|}{32 \pi \hat s E_e^{\rm cm}}
|\Delta_W|^2\\
&&\times
\displaystyle
\frac {2(s - m_b^2)(s - m_t^2)
m_t}{\sqrt{\hat s}} \sin\theta_{t}^{\rm cm} 
\left( (1 - \cos\theta_{t}^{\rm cm}) + \frac{m_b^2}{\hat s}
 (1+\cos\theta_{t}^{\rm cm})\right)
\end{array}
\end{equation}
and
\begin{equation}
\begin{array}{lcl}
\displaystyle
\sigma_{\rm unpol.}&=& \displaystyle \frac{g^4}{16}\int\,dx\,\overline
b(x)
\int d\cos\theta_{\overline t}^{\rm cm} \,
\frac{|\vec p_{\,\overline t}^{\,\rm cm}|}{32 \pi \hat s E_e^{\rm cm}}
|\Delta_W|^2\\&& \times 
(\hat s - m_b^2)(\hat s - m_t^2)
\displaystyle
\left[ \left((1 - \cos\theta_{t}^{\rm cm})^2 + \frac{m_b^2}{\hat s} 
\sin^2\theta_{t}^{\rm cm}\right)\right. \\
&& + 
\displaystyle \left.
\frac{m_t^2}{\hat s} \left(\sin^2\theta_{t}^{\rm cm} +
\frac{m_b^2}{\hat s}
 (1+\cos\theta_{t}^{\rm cm})^2\right)\right]
\end{array}
\end{equation}
The transverse polarization asymmetry, as anticipated, is proportional to 
Im~$f_{2L}$.

In Case (iii), the transverse polarization arising from Im~$f_{2R}$
vanishes because $\rho_{12}$ and $\rho_{21}$ are real, and the imaginary
part of $\rho_{12}$ appearing in the asymmetry is zero. 
In Case (i), The transverse polarization
due to Im~$f_{1R}$ is non-vanishing. However, it is found to be
proportional to $m_b/\sqrt{\hat s}$, and therefore negligibly small.
Thus, the measurement of a
non-vanishing transverse polarization would signal specifically 
the presence of a nonzero coupling Im~$f_{2L}$.

We evaluate the relevant integrals numerically.
We use the values $m_t = 172.5$ GeV, $\sin^2\theta_W = 0.223$, where $\theta_W$ is
the electroweak mixing angle,  $m_W = 80.379$ GeV, and $\Gamma_W = 
2.085$ GeV for the $W$ boson width. We employ CTEQL1 parton 
distributions with $Q=m_t$ in the five-flavour scheme.
The result is that the transverse polarization is 
$0.175\times {\rm Im} f_{2L}$ for $E_e = 60$ GeV,
and $0.149\times {\rm Im} f_{2L}$ for $E_e = 150$ GeV.

To estimate how well the asymmetry $A_{\rm pol.}$ can measure Im~$f_{2L}$ 
for a  certain integrated luminosity $L$, we 
assume that the number of asymmetric
events $\sigma_{\rm pol.}L = A_{\rm pol.}\sigma_{\rm unpol.}L$ is larger than the  
$\sqrt{N_{\rm SM}}$,  
the statistical fluctuation in the SM events in the absence of 
Im~$f_{2L}$, thus
allowing the discrimination of the polarization arising from the
anomalous coupling from the statistical background at the 1-$\sigma$
level. This is expressed by the inequality
\begin{equation}
A_{\rm pol.} \sigma_{\rm SM} L > \sqrt{\sigma_{\rm SM} L}. 
\end{equation}
Thus, the limit on Im~$f_{2L}$ that can be placed at the 1-$\sigma$
level with luminosity $L$ is
\begin{equation}
{\rm Im}~f_{2L}^{\rm lim} = \frac{1}{\sqrt{\sigma_{\rm SM}L} A_1},
\end{equation}
where $A_1$ is the value of the asymmetry for ${\rm Im}~f_{2L}=1$.
Using this expression, the limits on the coupling for the two energies
considered here and with an integrated luminosity of 
$L = 100~{\rm fb}^{-1}$ suggested in 
\cite{CDR, LHeCStudyGroup:2012zhm, haosun} 
are given in Table 1.
\begin{table}[hbt]
\begin{centering}
\begin{tabular}{cccc}
\hline
$E_e$ (GeV) & $\sigma_{\rm SM}$ (pb)&$A_1$ & Im~$f_{2L}^{\rm lim}$\\ 
\hline
60 & 2.004 &  0.175  & $1.28 \times
10^{-2}$\\
150& 6.410 & 0.149  & $8.38 \times 10^{-3}$ \\
\hline
\end{tabular}
\caption{The transverse top polarization asymmetry 
$A_1$ for unit Im~$f_{2L}$
and the limits Im~$f_{2L}^{\rm lim}$ on Im~$f_{2L}$ for the two 
electron energies $E_e =
60$ GeV and $E_e = 150$ GeV, for an integrated
luminosity of 100 fb$^{-1}$.}
\end{centering}
\end{table}

\section{T-odd observable from angular variables}
We now consider a second T-odd observable which arises when the
$\overline t$ decay is taken into account.
If we include the top decay $\overline t \to \overline b 
\ell^- \overline \nu_{\ell}$, it is possible to construct T-odd
observables without explicit recourse to top spin. For example a T-odd variable which may be constructed 
is $\epsilon_{\mu\nu\alpha\beta}\ p_e^\mu\ p_P^\nu\
p_t^\alpha\ p_\ell^\beta$.
It is a Lorentz-invariant quantity, and could be evaluated in any
frame. For example, in the lab frame
it evaluates to
 $2 E_e  E_P |\vec p_{\overline t}|E_l \sin\theta_{\overline t}
\sin\theta_\ell \sin\phi_\ell $, with all quantities calculated in the
lab frame. 
However, as stated earlier, the charged-lepton angular distribution is
independent of the anomalous couplings in the decay vertex 
to first order \cite{theorem}. 
Hence, 
to isolate the effect of anomalous coupling Im~$f_{2L}$ in the
production process and to avoid the influence of anomalous couplings 
in decay, 
we consider only a combination of angular variables,
even though it is not Lorentz
invariant. So our second T-odd observable is the expectation value of 
\begin{equation}
O =  \sin\theta_{\overline t} \sin\theta_\ell \sin\phi_\ell , 
\end{equation}
all variables being measured in the lab frame.

\comment{
The contribution to the expectation value, as mentioned earlier, comes
only from the imaginary parts of the couplings. Thus, only the
off-diagonal elements of $\rho$ are relevant, the diagonal elements
being real. 
The contributions of
$f_{2L}$ to $\rho_{12}$ and $\rho_{21}$ are:
      \begin{equation}
\rho_{12} =
\frac{g^4}{16}|\Delta_W|^2
  F_{2L}\frac{(\hat s-m_b^2)(\hat s-m_t^2)
           m_t}{\sqrt{\hat s}}\sin\theta_{t}^{\rm cm}
 \left[2-(1- m_b^2/{\hat s}) (1+\cos\theta_{t}^{\rm cm}) \right]
\end{equation}
      \begin{equation}\rho_{21} =
\frac{g^4}{16}|\Delta_W|^2
F_{2L}^* \frac{(\hat s-m_b^2)(\hat s-m_t^2)
          m_t}{\sqrt{\hat s}}\sin\theta_{t}^{\rm cm}
\left[ 2-(1-m_b^2/{\hat s}) (1+\cos\theta_{t}^{\rm cm}) \right]
\end{equation}
where
 \begin{equation}
F_{2L} = 1 + ({\rm Re} f_{2L} + i{\rm Im} f_{2L})\frac{(\hat s-m_t^2)}
{m_tm_W}.
\end{equation}
} 

The density matrix elements for the decay $\overline t \to \overline 
b \ell^- \overline \nu_\ell$, 
with an integration over all variables except the charged-lepton 
angular variables $\theta_\ell^0$ and $\phi_\ell^0$ in the top
rest frame, using the narrow-width approximation for the $W$, are (see
for example \cite{Godbole:2002qu})
\begin{equation}
\begin{array}{ccl}
        \Gamma_{11}& =& K(1 + \cos\theta_\ell^0)\\
        \Gamma_{22}& =& K(1 - \cos\theta_\ell^0)\\
        \Gamma_{12}& =& K\sin\theta_{\ell}^0 e^{i \phi_\ell^0}\\
        \Gamma_{21}& =& K\sin\theta_{\ell}^0 e^{-i \phi_\ell^0}
\end{array}
\end{equation}
where 
\begin{equation}
        K = g^4(m_t^2 - 2p_t\cdot p_\ell).
\end{equation}

In principle, the anomalous $tbW$ couplings would contribute to the
decay density matrix. However, restricting to linear order, the effect
of the
anomalous couplings on the normalized angular distributions has been 
shown to be absent \cite{theorem}. We therefore do not include anomalous
couplings, assuming them to be small enough to ignore the quadratic
powers. 

The cross section for the process in the lab frame is given
by 
 \begin{equation}
\begin{array}{rcl}
 \sigma& =& \displaystyle \int dx\, \overline b(x)
\int d\cos\theta_{\overline t}^{\rm lab} \int d\cos\theta_{\ell^-}^{\rm
lab}\int d\phi_{\ell^-}^{\rm lab}
\int dE_{\ell^-}^{\rm lab} 
 \frac{E_{\ell}^{\rm lab} p_{\overline t}^{\rm
cm}}{\hat s^{3/2}} 
\\ && \times\displaystyle 
 \frac{1}{512 (2\pi)^4 \Gamma_t\Gamma_Wm_tm_W}
\displaystyle
 (\rho_{11}\Gamma_{11}+\rho_{22}\Gamma_{22} + 
\rho_{12}\Gamma_{12} + \rho_{21}\Gamma_{21}).
	\end{array}
 \end{equation}
Here, $\Gamma_t$ ($\Gamma_W$) are the total widths of the $t$ (
$W$), $\overline b(x)$ is the $\overline b$ parton distribution 
function
in the proton, $E_\ell^{\rm lab}$ is the energy of the decay lepton in
the lab frame, and $p_{\overline t}^{\rm cm}$ is the magnitude of the
$\overline t$ three-momentum  in the $p_{\overline t}\nu_e$ cm frame.  
The expectation value of $O$ is then 
 \begin{equation}\label{expO}
\begin{array}{rcl}
 \langle O \rangle & =& \displaystyle \frac{1}{\sigma}\int dx\, \overline b(x)
\int d\cos\theta_{\overline t}^{\rm lab} \int d\cos\theta_{\ell^-}^{\rm
lab}\int d\phi_{\ell^-}^{\rm lab}
\int dE_{\ell^-}^{\rm lab}
 \frac{E_{\ell}^{\rm lab} p_{\overline t}^{\rm
cm}}{\hat s^{3/2}} 
\\ && \times\displaystyle 
 \frac{1}{512 (2\pi)^4 \Gamma_t\Gamma_Wm_tm_W}
 (\rho_{11}\Gamma_{11}+\rho_{22}\Gamma_{22} + 
\rho_{12}\Gamma_{12} + \rho_{21}\Gamma_{21})\,O.
	\end{array}
 \end{equation}

The contribution to the expectation value, as mentioned earlier, comes
only from the imaginary parts of the couplings. 
Thus, only the off-diagonal
elements of $\rho$ are relevant, the diagonal elements being real.
As in the case of the transverse polarization, it turns out the
observable $\langle O \rangle $  is zero in Case (iii) of nonzero
anomalous 
coupling $f_{2R}$, because in that case $\rho_{12}$ and $\rho_{21}$ are
real. 
\comment{
\begin{equation}
\begin{array}{rcl}
      \rho_{12}\;\, = \;\,
	\rho_{21} &=&
\displaystyle
\frac{g^4}{16}|\Delta_W|^2 
\frac{\hat s- m_b^2)(\hat s - m_t^2)m_t}{\sqrt{\hat s}}  
     \sin\theta_{\overline t}^{\rm cm}[(1 - \cos\theta_{\overline t}^{\rm} )\\&&
\displaystyle
          + \frac{m_b^2}{\hat s}|F_{2R}|^2  (1 +
     	\cos\theta_{\overline t}^{\rm cm})],
\end{array}
\end{equation}
where
 \begin{equation}	
F_{2R} = 1 + ({\rm Re} f_{2R} + i{\rm Im} f_{2R})\frac{(\hat s-m_b^2)}
{m_bm_W}.
\end{equation}
} 
There is a nonzero contribution in Case (i) with nonzero $f_{1R}$, 
but it turns out to be vanishingly small, the contributions of the
$f_{1R}$ to $\rho_{12}$ and $\rho_{21}$ being proportional 
to $m_b/\sqrt{\hat s}$.
\comment{
\begin{equation}
\begin{array}{rcl}
\rho_{12} = \rho_{21}^* \!\!\!\! &=& \!\!\! 
\displaystyle
\frac{g^4}{16}|\Delta_W|^2 
(\hat s - m_b^2) (\hat  s - m_t^2) \sin\theta_{\overline t}^{\rm cm} [ (m_b^2/\hat s)
(m_t/\sqrt{\hat s}) (1 + \cos\theta_{\overline t}^{\rm cm}) \\&&  - (m_b/\sqrt{\hat s}) 
({\rm Re} f_{1R} + i {\rm Im} f_{1R}) 
+ (m_t/\sqrt{\hat s})(1-\cos\theta_{\overline t}^{\rm cm})].
\end{array}
\end{equation}
\begin{equation}
\begin{array}{rcl}
\rho_{21} &&= 
\displaystyle
\frac{g^4}{16}|\Delta_W|^2 
(\hat s - m_b^2) (\hat  s - m_t^2) \sin\theta_{\overline t}^{\rm cm} [ (m_b^2/\hat s)
(m_t/\sqrt{\hat s}) (1 + \cos\theta_{\overline t}^{\rm cm})\\&&   - (m_b/\sqrt{\hat s}) 
({\rm Re} f_{1R} - i {\rm Im} f_{1R}) 
+ (m_t/\sqrt{\hat s})(1-\cos\theta_{\overline t}^{\rm cm})]
\end{array}
\end{equation}
} 
We are thus left with Case (ii), where, as seen earlier, Im~$\rho_{12}$
is nonzero. We therefore evaluate the expectation value of $O$ for Case
(ii), using eq. (\ref{expO}). 

For the observable $O$ to be measurable at the 1-$\sigma$ level with an
integrated luminosity $L$, its
expectation value should satisfy
\begin{equation}
\vert \langle O \rangle -  \langle O \rangle_{\rm SM} \vert 
> \frac{\sqrt{ \langle O^2 \rangle_{\rm SM} - 
\langle O \rangle_{\rm SM}^2}}{\sqrt{\sigma_{\rm SM} L}},
\end{equation}
where $\langle \dots \rangle$ denotes the expectation value.
For our case, $\langle O \rangle_{\rm SM} = 0$. 
The 1-$\sigma$ limit on the coupling Im~$f_{2L}$ is then given by 
\begin{equation}
{\rm Im}~f_{2L}^{\rm lim} =
\frac{\sqrt{ \langle O^2 \rangle_{\rm SM}  
}}{\left(\langle O \rangle_{{\rm Im} f_{2L} = 1}
\right)\sqrt{\sigma_{\rm SM} L}}.
\end{equation}
One may remark that the transverse polarization considered above may be
thought of as the expectation value of an observable 
\begin{equation}
O = \epsilon(
\vec p_{e^-} \times \vec
p_{\overline t} \cdot \vec s_{\overline t} ),
\end{equation}
 where $\epsilon$ is the
usual sign function taking values $+1$ or $-1$ when the argument is
respectively positive and negative. Then, $O^2 = 1$, $\langle O \rangle
$ reduces to the asymmetry $A_{\rm pol.}$, and the above 
expression for the limit reduces to the one given above 
for the limit from measurement of the transverse polarization asymmetry.

The expectation values
and the limits on Im~$f_{2L}$ for the two electron energies $E_e =
60$ GeV and $E_e = 150$ GeV are given in Table 2, assuming an integrated
luminosity of 100 fb$^{-1}$.
The limits in Table 2 obtainable from the observable $O$ are comparable
to those in Table 1 which would result from the transverse polarization,
though somewhat better. A more detailed study of the kinematics and
relevant cuts for particle identification would be required to get more
accurate sensitivities. However, our numbers should give a reasonable 
indication of the true values.

\begin{table}[htb]
\begin{centering}
\begin{tabular}{ccccc}
\hline
$E_e$ (GeV) & $\sigma_{\rm SM}$ (pb)&$\langle O \rangle_1$  
&$\langle O^2 \rangle_{\rm SM}$ & Im~$f_{2L}^{\rm lim}$\\ 
\hline
60 & 0.231 &  $4.89\times 10^{-2}$ &$5.39\times 10^{-3}$ & $9.88 \times
10^{-3}$\\
150& 0.737 & $1.83\times 10^{-1}$ & $7.16 \times 10^{-2} $ & $5.39 \times 10^{-3}$\\
\hline
\end{tabular}
\caption{The SM cross section for $\overline t$ production and decay
into the leptonic channel, the expectation value of $O$ for unit 
coupling, the expectation value of $O^2$ in
the SM, 
and the limits Im~$f_{2L}^{\rm lim}$ on  Im~$f_{2L}$for the two electron energies $E_e =
60$ GeV and $E_e = 150$ GeV, for an integrated
luminosity of 100 fb$^{-1}$.}
\end{centering}
\end{table}
 
\section{Conclusions and Discussion}

We have considered two typical T-odd observables which might be measured
in the process $e^- p \to \nu_e \overline t X$ in an extension of the
SM with anomalous $tbW$ interactions. These T-odd observables are
expected to be dependent on the imaginary part of the anomalous
coupling.

Of the three anomalous couplings $f_{1R}$, $f_{2L}$ and $f_{2R}$, the
imaginary parts of the first two give rise to a nonzero
transverse polarization for the $\overline t$. Im~$f_{2R}$ gives a
vanishing transverse polarization. The transverse polarization arising
from $f_{1R}$ is negligibly small, being proportional to $m_b/\sqrt{\hat
s}$. Similarly, the expectation value of $O$ turns out to be nonzero
and appreciable only in the case of the coupling Im~$f_{2L}$. 
Thus, a measurement of the two T-odd observables we suggest here would
serve to isolate the coupling Im~$f_{2L}$ from amongst the three
anomalous couplings available.

The 1-$\sigma$ limits on Im~$f_{2L}$ we derive are of the order of 
10$^{-2}$, with either T-odd variable. The limit is better with
higher electron energy. In going from $E_e = 60$ GeV to $E_e =
150$ GeV, improvement is by a factor of less than 2.
Our calculation is restricted to a single channel for the 
antitop decay. But it could be extended to the two leptonic channels as
well.  In that case,
the sensitivity would be expected to be better by approximately a factor
 of $\sqrt{3}$.
It is conceivable that the sensitivity would be better at colliders
planned to operate at even higher cm energies, though these have not
been considered here.
 
The sensitivities for the measurement of Im~$f_{2L}$ estimated here
would need refinement.
A more reliable estimate would require a detailed
study of the detector efficiencies and imposing of realistic kinematic
cuts for identification of the particles like the $\overline t$, 
which is not attempted here.
Such a study would be worthwhile to carry out.

One may construct other more complicated T-odd observables, which are
products of the observable considered above with kinematic constructs
which are T-even, so that the product is odd under T. It may be possible
to optimize the sensitivity with the use of such a combination.

We do not consider hadronic top decays, which have a larger probability,
 since it would be difficult, if
not impossible, to measure a triple product of momenta, which would
require measurement of the charges of the quarks. However, given the
advantage of an enhanced sample of events in such a case, it would be
worthwhile to attempt using hadronic decays as well.

We have discussed naive T violation, not necessarily accompanied by CP
violation. However, if there is a possibility of employing $e^+$ beams
in addition to $e^-$ beams, thereby measuring also 
the $e^+p \to t X$ process, it would be possible to compare cross
sections for the two conjugate processes, thereby enabling observation
of CP violation, if it exists.

The possibility of a polarized $e^-$ beam has been discussed in the
context of an $ep$ collider. Since the coupling of the electron to the
$W$ is left-handed, a left-handed polarization of the $e^-$ beam would
increase the cross section, and make the process more sensitive to the
extent of the degree of left-handed polarization.

\noindent {\bf Acknowledgement} 
The work was supported by the
Senior Scientist programme of the Indian National Science Academy, New
Delhi. The author thanks the referee for suggesting useful improvements in 
the mansucript.

	\end{document}